\documentclass[traditabstract]{aa} % for the abstract without structuration 
\usepackage{graphicx,caption}
\usepackage{marginnote}
\usepackage{txfonts}
\usepackage{subfigure}
\usepackage{natbib}
\usepackage{amstext}
\usepackage[verbose]{placeins}
\usepackage{tikz}
\usepackage{array}
\usepackage{amssymb}
\usepackage{xcolor}
\usepackage{hyperref}

\newcommand{\bd}{\begin{displaymath}}
\newcommand{\ed}{\end{displaymath}}
\newcommand{\be}{\begin{equation}}
\newcommand{\ee}{\end{equation}}
\newcommand{\beaa}{\begin{eqnarray*}}
\newcommand{\eeaa}{\end{eqnarray*}}
\newcommand{\bea}{\begin{eqnarray}}
\newcommand{\eea}{\end{eqnarray}}

\newcommand*{\figuretitle}[1]{%
  {\bigskip
    \centering%   <--------  will only affect the title because of the grouping (by the
    {\textbf{#1}}%              braces before \centering and behind \medskip). If you remove
    \par\medskip}%            these braces the whole body of a {figure} env will be centered. % can use \smallskip and \large before {\textbf{}}
}

\begin{document}

   \title{Photometric redshift estimation with a convolutional neural network: NetZ}

  \titlerunning{Photometric Redshift with an CNN}

   \author{S. Schuldt\inst{1}\inst{,2}
          \and
          S. H. Suyu\inst{1}\inst{,2}\inst{,3}
          \and
          R. Ca\~{n}ameras\inst{1}
          \and
          S. Taubenberger\inst{1}
          \and
          T. Meinhardt\inst{4}
          \and
          L. Leal-Taix\'{e}\inst{4}
          \and
          B.C. Hsieh\inst{3}
          }

   \institute{Max-Planck-Institut f\"ur Astrophysik, Karl-Schwarzschild Str.~1, 85741 Garching, Germany \\
              \email{schuldt@mpa-garching.mpg.de}
         \and  
             Physik Department, Technische Universit\"at M\"unchen, James-Franck Str. 1, 85741 Garching, Germany
         \and
             Institute of Astronomy and Astrophysics, Academia Sinica, 11F of ASMAB, No.1, Section 4, Roosevelt Road, Taipei 10617, Taiwan
         \and
            Informatik Department, Technische Universit\"at M\"unchen, Bolzmannstr. 3, 85741 Garching, Germany
             }

   \date{Received --; accepted --}

% \abstract{}{}{}{}{} 
   % 5 {} token are mandatory

  \abstract
  % context heading (optional)
  % {} leave it empty if necessary  
      {
        Galaxy redshifts are a key characteristic for nearly all extragalactic studies. Since spectroscopic redshifts
require additional telescope and human resources, millions of galaxies
are known without spectroscopic redshifts. Therefore, it is crucial to
have methods for estimating the redshift of a galaxy based on its
photometric properties, the so-called photo-$z$. We have developed NetZ, a
new method using a convolutional neural network (CNN) to predict the
photo-$z$ based on galaxy images, in contrast to previous methods that
often used only the integrated photometry of galaxies without their
images. We use data from the Hyper Suprime-Cam Subaru Strategic
Program (HSC SSP) in five different filters as the training data. The
network over the whole redshift range between 0 and 4 performs well
overall and especially in the high-$z$ range, where it fares better than other methods
on the same data. We obtained a precision $|z_\text{pred}-z_\text{ref}|$
of $\sigma = 0.12$ (68\% confidence interval) with a CNN working for
all galaxy types averaged over all galaxies in the redshift range of 0
to $\sim$4. We carried out a comparison with a network trained on point-like sources, highlighting the importance of morphological information for our redshift estimation. By limiting the scope to smaller redshift ranges or to luminous red
galaxies (LRGs), we find a further notable
improvement. We have published more than 34 million new photo-$z$ values
predicted with NetZ.
This
shows that the new method is very simple and swift in application, and,
importantly, it covers a wide redshift range that is limited only by the
available training data. It is broadly applicable, particularly with regard to upcoming surveys such as the Rubin
Observatory Legacy Survey of Space and Time, which will provide
images of billions of galaxies with similar image quality as HSC. Our HSC photo-z estimates are also beneficial to the Euclid survey, given the overlap in the footprints of the HSC and Euclid.
%$^*$ The catalog is available at: \url{https://www.dropbox.com/sh/grjfo0gkcxsj9n2/AAD-B7D6m7_1i6GGTX0Ionwja?dl=0}.
}
  % aims heading (mandatory)
   %{}
  % methods heading (mandatory)
   %{}
  % results heading (mandatory)
   %{}
  % conclusions heading (optional), leave it empty if necessary 
   %{}

   \maketitle

\section{Introduction}
\label{sec:introduction}

Past imaging surveys have detected billions of galaxies over the sky,
a number that will grow substantially with forthcoming wide-field
surveys, such as the Rubin Observatory Legacy Survey of Space and Time (LSST). In most applications for which galaxies are
used, redshifts are needed, but spectroscopic redshifts are available
only for a small fraction of them. Therefore photometric redshift
techniques \citep[hereafter photo-$z$, see][ and
references therein]{hildebrandt10} were developed and improved over the last decades
\citep[e.g.,][]{coupon09, hildebrandt08, hildebrandt12, dahlen13,
  bonnett16, tanaka18}. Typically, photometry in multiple wavelength
bands has been used to minimize the difference between spectroscopically
confirmed redshifts and the predicted photometric redshifts.

Today, there are two main families of photo-$z$ methods, namely:\ template
fitting and machine learning (ML) methods. They are complementary to one another and
both are capable of predicting very precise photo-$z$. Template fitting
codes \citep[e.g.,][]{arnouts99, bolzonella00, feldmann06, brammer08, duncan18b}
are mainly based on galaxy spectral energy distribution (SED)
template libraries. This method is physically motivated and
well studied thus far. The templates are used to match the observed colors with
the predicted ones (via the so-called nearest neighbor
algorithms). Such an approach represents the opportunity to provide
photo-$z$ estimates in regions of color-magnitude space where no
reference redshifts are available.
Additionally, ML provides another approach to get very precise and fast photo-$z$
estimates \citep[e.g.,][]{tagliaferri03, collister04, lima08, wolf09,
  carliles10, singal11, hoyle16, tanaka18, bonett15, DIsario18, eriksen20, schmidt20}. The main requirement is a
training sample with known (i.e., spectroscopic or very good photo-$z$)
reference redshifts, which should match the expected redshift
distribution. Depending on the network architecture, ML codes generally look for specific features in the training sample and try to
extract the important information. So far, most algorithms are based on
photometric parameters like color-magnitude measurements or also
size-compactness measurements and often limited to a narrow redshift
range, for example, up to $z = 1$ \citep[e.g.,][]{bonett15, hoyle16, sadeh16, almosallam16, pasquet18,   pasquet19, eriksen20, campagne20}.

Based upon the success of CNNs in image processing, we move on to our investigation of a network
that estimates photo-$z$ based directly on images of galaxies. This is
similar to the work done by \citet{hoyle16}, where images of galaxies
are converted into magnitude images and pixel color maps to feed the
architecture, however, our network accepts the images directly as
observed. Moreover, while \citet{hoyle16} used a classification network
whereby the galaxies are sorted into redshift bins between 0 and 1, we
use a regression network. This means our network predicts one specific
number for the galaxy redshift. 
% and was limited to the range between $0<z<1$. 
The work presented by \citet{DIsario18} and \citet{pasquet19} explores both networks with CNN layers on SDSS galaxies to obtain a probability density function (PDF). \citet{DIsario18}  tested networks for either quasars or galaxies as well as a combination of stars, quasars, and galaxies. For the galaxy sample, they limited their study to $0<z<1$, although most galaxies are at the lower end, such that \citet{pasquet19} directly limit the range up to $z=0.4$. In comparison to those two networks, we have many more galaxies with higher redshifts ($z \sim 1-3$) and thus we do not set limits on the redshift
range for the purpose of obtaining a more powerful network that
is directly applicable to the expected redshift range covered by
LSST. Based on the available reference redshifts, we tested the
performance up to a redshift of 4. Since we provide images of
different filters, our CNN is able to extract the color and magnitude
parameters internally and output a photo-$z$ value at the end. It is
trained on images observed in five different filters, specifically on Hyper Suprime-Cam Subaru Strategic
Program \citep[HSC SSP, hereafter HSC;][]{Aihara+18}
\textit{grizy} images of galaxies with known spectroscopic or reliable
$\sim$$30$-band photometric redshifts.

The outline of the paper is as follows. In Sect.~\ref{sec:data}, we describe the training data we applied and we give a short introduction and overview of the network architecture we used in Sect.~\ref{sec:network}. Our main network, NetZ$_\text{main}$, is presented in Sect.~\ref{sec:results} and we compare our results to other model techniques in Sect.~\ref{sec:comparison}. We show, in Sect.~\ref{sec:specific}, our results based on the network NetZ$_\text{LRG}$, which is specialized for Luminous Red Galaxies (LRGs) and  NetZ$_\text{lowz}$, which is specialized for the low redshift range. We summarize our results in Sect.~\ref{sec:summary}.

\FloatBarrier
\section{Training data}
\label{sec:data}

We use images from PDR2 of the HSC-SSP\footnote{HSC webpage:
  https://hsc-release.mtk.nao.ac.jp/doc/} survey \citep{aihara19}
for the training of the CNN. The HSC is a wide-field
optical camera with a field of view of 1.8 square degrees (1.5 degree
in diameter) installed at the 8.2m Subaru Telescope. The data release
covers over 300 square degrees of the night sky in five optical
filters known as \textit{grizy}. The exposure time is 10 minutes for the
filters \textit{g} and \textit{r} and 20 minutes for \textit{i},
\textit{z}, and \textit{y}, yielding limiting magnitudes of around
26. The pixel size is $0.168\arcsec$, such that our cutouts with
$64\times64$ pixels result in images of around
$10\arcsec\times10\arcsec$. The median seeing in the i-band is
0.6$\arcsec$.

\begin{figure}[ht!]
  \centering
\includegraphics[angle = 0, trim= 0 250 0 250, clip, width=\columnwidth]{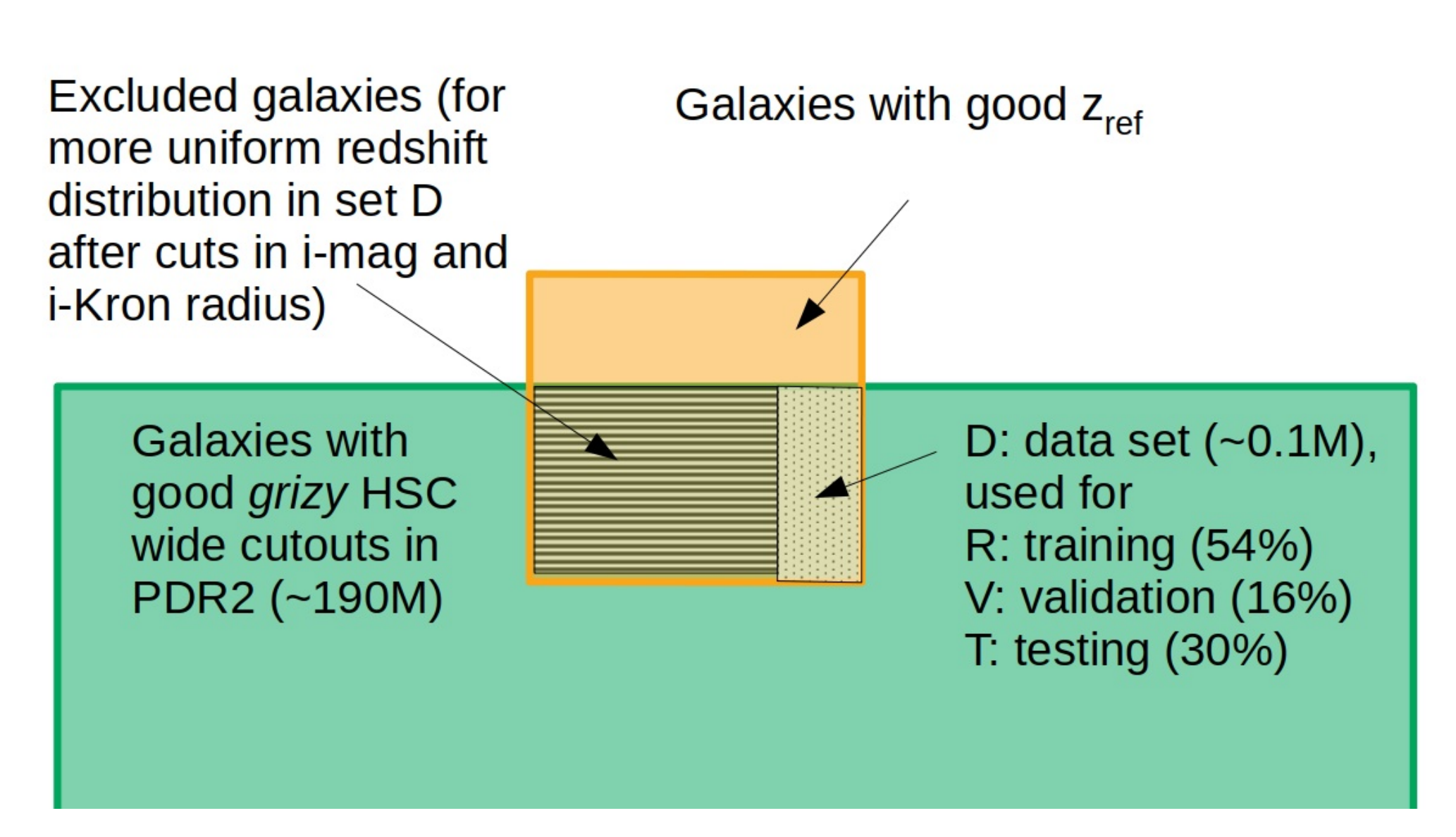}
\caption{Sketch of the available data and the intersection of the data
  D (dotted) used for training (R), validation (V), and testing (T) of
  the main network NetZ$_\text{main}$, as presented in Sect.~\ref{sec:results}.}
\label{fig:SampleSketch}
\end{figure}

The catalog of all available galaxies from HSC PDR2 in the wide area that pass the following criteria:\ \ \ 
{
\begin{description}
%\item[$\bullet$] \{$grizy$\}\_blendedness\_flag is \textit{True}
%\item[$\bullet$] \{$grizy$\}\_blendedness\_flag\_nocentroid is \textit{True}
% \item[$\bullet$] \{$grizy$\}\_blendedness\_flag\_noshape is \textit{True}
% \item[$\bullet$] \{$grizy$\}\_pixelflags\_suspectcenter is \textit{True}

\item[$\bullet$] \{$grizy$\}\_cmodel\_flux\_flag is \textit{False}
\item[$\bullet$] \{$grizy$\}\_pixelflags\_edge is \textit{False}
\item[$\bullet$] \{$grizy$\}\_pixelflags\_interpolatedcenter is \textit{False}
\item[$\bullet$] \{$grizy$\}\_pixelflags\_saturatedcenter is \textit{False}
\item[$\bullet$] \{$grizy$\}\_pixelflags\_crcenter is \textit{False}
\item[$\bullet$] \{$grizy$\}\_pixelflags\_bad is \textit{False}
\item[$\bullet$] \{$grizy$\}\_sdsscentroid\_flag is \textit{False}
\end{description}
}
\noindent includes around 190 Million galaxies and is represented by a
green box in Figure \ref{fig:SampleSketch}.
The corresponding HSC images can be used as input data for the network NetZ.

As ground truth, we use the spectroscopic redshifts provided by the HSC
team, which is a collection from various spectroscopic surveys
(zCOSMOS DR3 \citep{lilly09}, UDSz \citep{bradshaw13, mclure12},
3D-HST \citep{skelton14, momcheva16}, VVDS \citep{lefevre13}, VIPERS
PDR1 \citep{garilli14}, SDSS DR14 \citep{alam15}, GAMA DR2
\citep{liske15}, DEEP3 \citep{davis03, newman13}, PRIMUS DR1
\citep{coil11, cool13}). %{\bf Since we aim to obtain a network that is applicable to all morphological types, we use spec-z from surveys that are not pre-selecting objects based on morphology.  Specifically, we exclude objects from SDSS   BOSS/eBOSS as those surveys are targeting specifically LRGs as well as known quasars. Furthermore, we do not include WiggleZ \citep{drinkwater10} since this catalog is biased towards UV bright emission line galaxies and therefore excludes passive galaxies. The sample FMOS-COSMOS \citep{silverman15} mostly targeted star-forming galaxies with $1.4 < z < 1.7$ and $M_\text{stellar} > 10^{10} M_\text{sun}$, and got ~400 redshifts. Although this is a relative small sample and would not dominate our whole data set at the end, it is excluded in our data set as it focuses on star-forming galaxy rather than a generic galaxy type. With this collection of spectroscopic confirmed redshift catalog, we obtain a magnitude-limited sample without morphological pre-selection. This spec-$z$ sample is cleaned with the following criteria}:
Since we aim to obtain a network that is applicable to all
morphological types, the above list does not include spectroscopic
surveys that are most strongly biased towards specific galaxy types of similar
morphology. Specifically, we do not consider objects from SDSS
BOSS/eBOSS to train our main network NetZ$_\text{main}$ as those surveys
explicitly target LRGs at z<1 and known quasars. We do consider
training exclusively on LRGs in our separate network NetZ$_{\text{LRG}}$.
Furthermore, we do not include the WiggleZ catalog \citep{drinkwater10},
which targets UV bright emission line galaxies and which would further
steepen the redshift distribution of the training set at low-redshift
(see below). Despite unavoidable biases due to the selection function of
each survey, we expect that this collection
of spectroscopically-confirmed redshifts has limited morphological
pre-selection. This spec-$z$ sample is cleaned with the following criteria:

\begin{description}
 \item[$\bullet$] source type is \textit{GALAXY} or \textit{LRG}
 %\item[$\bullet$] Kron radius in i band is below $5\arcsec$ because of image cutout size
 \item[$\bullet$] $z>0$
 \item[$\bullet$] $z \ne 9.9999$\footnote{This is the upper limit of the catalog and thus treated as no spec-$z$ available, i.e. excluded}
 \item[$\bullet$] $0<z_{\text{err}}<1$
 \item[$\bullet$] the galaxy identification number (\textit{ID}) is unique
 \item[$\bullet$] specz\_flag\_homogeneous is \textit{False} (homogenized spec-$z$ flag from HSC team)
\end{description}

This spec-$z$ sample is used in combination with COSMOS2015
\citep{laigle18}, a photo-$z$ catalog of the COSMOS field based on
around 30 available filters, where we enforce the following criteria:

\begin{description}
\item[$\bullet$] flag\_capak is \textit{False}
\item[$\bullet$] type = 0 (only galaxies)
\item[$\bullet$] $\chi^2$(gal) $< \chi^2$(star) and $\chi^2$(gal)/Nbands $< 5$ (fits are reasonable and better than stellar alternatives)
\item[$\bullet$] ZP\_2$ < 0$ (no secondary peak)
\item[$\bullet$] log$( M_\star)  > 7.5$ (stellar mass successfully recovered)
\item[$\bullet$] $0<z<9$ 
\item[$\bullet$] max($z_{84} - z_{50},z_{50} - z_{16}) < 0.05(1+z)~ (1 \sigma$-redshift dispersion $<5$\%)
\end{description}

This selection primarily follows  the criteria from the other HSC photo-$z$ methods \citep{tanaka18, nishizawa20}. We then select galaxies with $i$-band magnitudes brighter than 25 mag and a
Kron radius larger than $0.8\arcsec$ in the $i$ band. The limit on the
Kron radius is chosen with the aim of obtaining a set that best represents the
sample that we are applying NetZ to. These criteria ensure that we have accurate and reliable reference redshifts for our training, validation, and testing. While such criteria could lead to potential selection bias in the objects, our combination of photo-$z$ and spec-$z$ helps mitigate selection biases.
Furthermore, we verify that the color space spanned by the objects from the cleaned catalog is similar to that of the objects in the HSC PDR2 with a Kron radius above 0.8”.  This allows us to apply the trained NetZ based on the reference redshifts to those HSC PDR2 galaxies. The cleaned catalog used for training, validation, and testing is shown as a yellow box in Figure \ref{fig:SampleSketch}, and the overlap with available good HSC images in all five filters (green box) contains 406,540 galaxies.

\begin{figure}[ht!]
  \centering
\includegraphics[angle = 0, trim=0 0 0 0, clip, width=1.0\columnwidth]{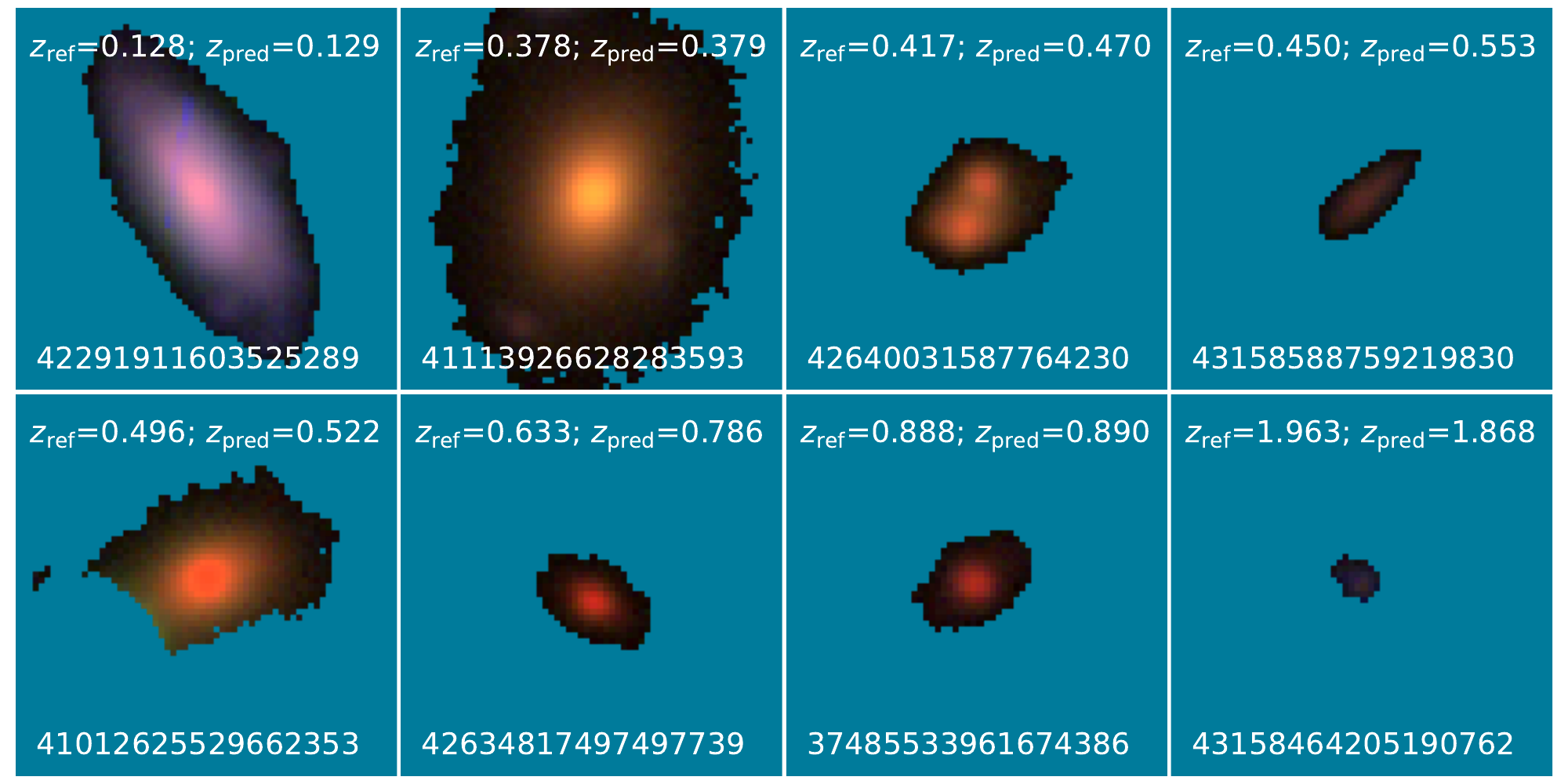}
% this plot is based on test set of Network_0340 as NetZ_main
\caption{Overview of galaxies from our data set. The masked neighbouring objects and background are shown in blue and have pixel values of zeros in the image. The images are $10.75\arcsec\times10.75\arcsec$ ($64\times 64$ pixels) and based on the three filters $g$, $r$, and $i$.  In each panel, the reference and predicted redshifts of the object are indicated at the top and the HSC identification number is at the bottom.}
\label{fig:mosaic}
\end{figure}

Based on various tests during the development stage, we found a significant
improvement by masking the background and surrounding objects next to
the galaxy of interest with the source extractor \citep{bertin96} before
feeding them into the CNN. As a boundary, we use the $3 \sigma$ level of
the background. Fully deblended neighboring objects in the field can be excluded by
requesting the object center to be within five pixels of the image
center. With this method, we keep only the central galaxy(ies) in the image
cutout. At the end we convolve the extracted image with a gaussian
kernel of size 3$\times$3 pixels and a width of 1.5 pixels to smooth
out the boundaries very slightly. We show color images of random galaxies from our NetZ$_\text{main}$ test sample in Figure~\ref{fig:mosaic} as examples. The masked background is shown in blue and has pixel values set to zeros in the image. We provide the reference redshift, which can be either a spectroscopic or photometric redshift, and our predicted redshift at the top of each image. The HSC identification number is given in the bottom of each image. From these examples, it can be seen that the extraction procedure works well overall, but has its limitations; for instance, the first image of the second row is partly truncated because of a masked bright neighbouring object. Since this procedure is aimed at masking only obvious and well deblended companions, while beeing purposely conservative and retaining blended galaxies. Therefore, the third image in the first row in Figure~\ref{fig:mosaic} is expected.
%For those galaxies passing our criteria, we find a better accuracy for the bright galaxies than fainter galaxies.

The reference redshift selection criteria described above give us a sample of galaxies with accurate reference redshifts, $z_\text{ref}$. Since the sample D is dominated by galaxies with
$z_\text{ref}<1$, we test the effect of data augmentation. Explicitly, we include
rotated images for $z_\text{ref}>1$, and in addition, mirrored images 
for $z_\text{ref}>2$.
An alternative to data augmentation is to introduce weights for the galaxies. For example, \citet{lima08} proposed a relative weighting of galaxies in order to match their spectroscopic sample to observables of the photometric sample. Although we could also adapt a similar weighting scheme to balance the redshift distribution, we favor the data augmentation technique that is commonly used in neural networks.

\begin{figure}[ht!]
  \centering
\includegraphics[angle = 0, trim=0 0 0 0, clip, width=1.0\columnwidth]{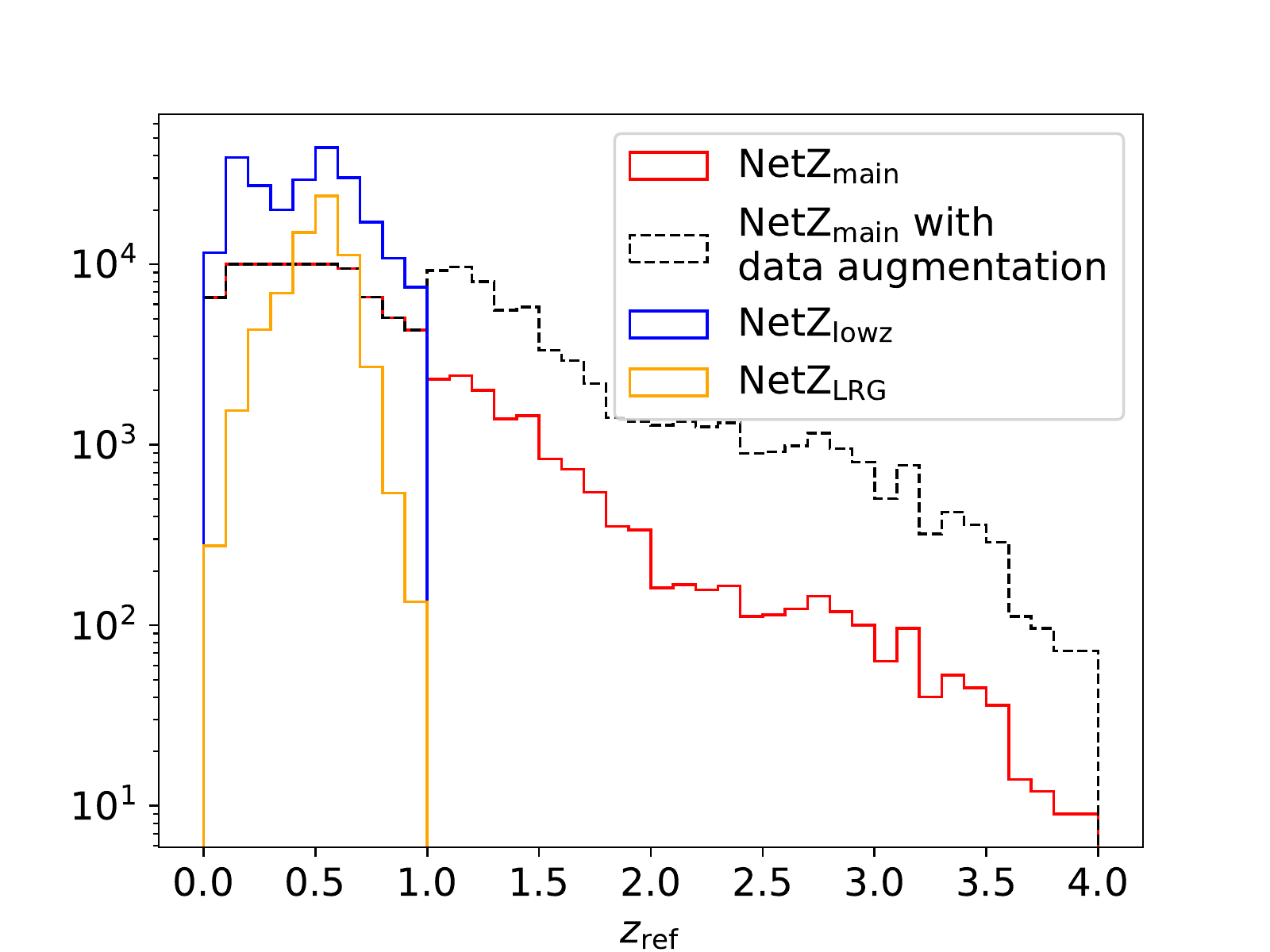}
% this plot is based on test set of Network_0340 as NetZ_main
\caption{Histograms of the redshift samples used in this work. For NetZ$_\text{main}$, we show the original redshift distribution in red, and the data augmented distribution in dashed black (with more galaxies at $z_{\rm ref}>1$) that was used for our final network. The distribution used to train our two specialized networks (see Sect.~\ref{sec:specific} for details) is overplotted for NetZ$_\text{lowz}$ in blue and for Net$Z_\text{LRG}$ in orange.}
\label{fig:hist_zref}
\end{figure}

Since the distribution of the reference redshifts in the training set is very
important for the network and still dominated by the lower redshift end, we limit each redshift bin of width 0.01 to
have no more than 1000 galaxies from those passing the above criteria. With this limit, we obtain a uniform distribution up to $z_\text{ref} \sim 1.5$
This essentially limits the number of low-redshift galaxies that
would otherwise be over-represented in the training set. As a result,
the redshift distribution becomes more uniform and allows the CNN to
learn and predict redshifts for the full redshift range rather than only
the lower redshift end. The excluded galaxy sample is marked in the
underlying yellow box with lines in Figure~\ref{fig:SampleSketch}, while sample D is used for our
main network NetZ$_\text{main}$, shown with a red histogram in Figure~\ref{fig:hist_zref}. We show the distribution of the augmented sample as a black dashed histogram in Figure~\ref{fig:hist_zref}.

\FloatBarrier
\section{Deep learning and the network architecture}
\label{sec:network}

Neural networks (NN) are very powerful tools that serve many different tasks, especially in works involving a huge amount of data. Substantial efforts
have therefore been dedicated to deep learning (DL) developments in
recent years. In general, for supervised learning, it is necessary to have a data set
where the input and output, that is, the so-called ground truth, are known. On
this data, the network is trained and can afterwards be applied to new
data where the output is not known. The main advantages of NN include the
variety of architectures and thus the broad range of problem they can
be applied to, as well as the speed of those networks in comparison to
other methods. Generally, there are two kinds of networks:
classification networks distinguish between different classes of
objects, whereas regression networks predict specific numerical
quantities. The latter is the kind of network we are using here, namely
it is the network that predicts a specific value for  the redshift of a galaxy.

\begin{figure*}[ht!]
\centering
\includegraphics[angle = 0, trim=1 0 0 0, clip, width=2.0\columnwidth]{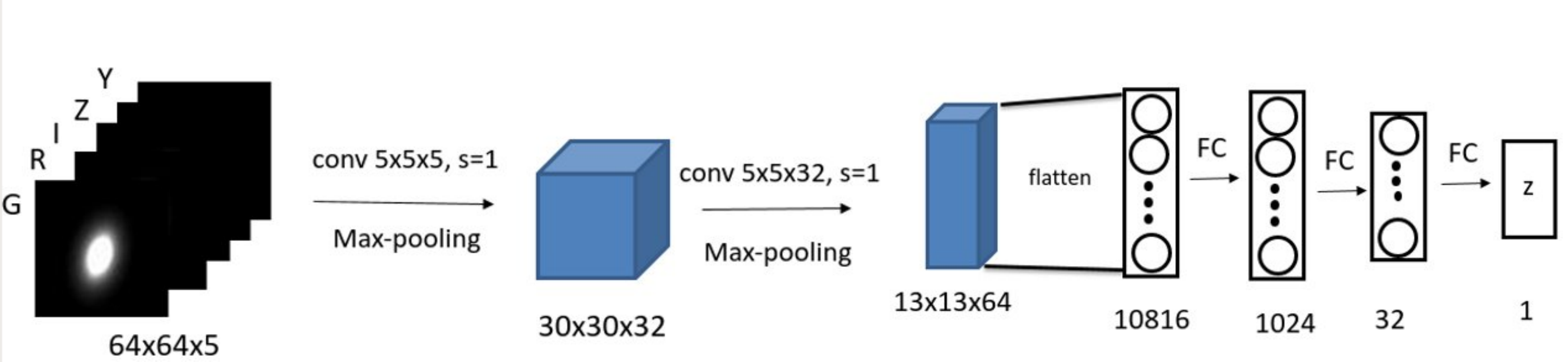}
\caption{Overview of the CNN architecture. It contains two
  convolutional (conv) layers with max pooling and three fully
  connected (FC) layers. The input consists of images of size 64$\times$64
  pixels in five different filters ($grizy$). The output displays the predicted
  photometric redshift.}
\label{fig:CNNOverview}
\end{figure*}

Depending on the task, there are different types of networks. Since
our input consists of images of galaxies, a typical type is the CNN
where the fully-connected (FC) layers are preceded by a number of
convolutional (conv) layers. The detailed architecture depends on
various parameters such the specific task, the size of the images,
and the size of the data set. We tested different architectures and
found an overall good network behavior with two convolutional layers
followed by three FC layers. We tested different constructions of CNN
architecture by varying the number of convolutional or FC layers, strides, and kernel
sizes but with no improvement. A sketch of the final architecture is shown in Figure
\ref{fig:CNNOverview}. The input consists of five different filters
for each galaxy and each image has a size of 64 $\times$ 64 pixels,
corresponding to an image size of around $10 \arcsec \times 10
\arcsec$. The convolutional layers have stride $s=1$ and a kernel size
of 5 $\times$ 5 $\times$ $C$, where $C=5$ in the first convolutional
layer, and $C=32$ in the second layer. We used 32 kernels and 64
kernels in the first and second convolutional layers,
respectively. Each convolutional layer is followed by max pooling of
size 2 $\times$ 2 and of stride $2$. This results in a data cube of
size $13 \times$ 13 $\times$ 64, which, after flattening, is passed on to
the FC layers to obtain the single output value, namely, the redshift of the
galaxy.

Independent of the network architecture, the network can contain
hundreds of thousands (or more) neurons. Even though at the beginning, the
values of the weight parameters and bias of each neuron are random,
they are updated at every iteration of the training. To see the
network performance after the training, we need to split the data into three
sets, the training set R, the validation set V, and the test set T (see
Figure \ref{fig:SampleSketch}). In our case, we used 56\% of the data
set as training set, 14\% as validation set, and 30\% as test set.  We
trained over 300 epochs and divided each epoch into a number of
iterations by splitting the training, validation, and test set into
batches of a size $N$. In each iteration, a batch is passed through the
entire network to predict the redshifts $z_\text{pred}$ (forward propagation). The
difference between those predicted values and the ground truth is
quantified by the loss function $L$, where we use the mean-square-error
(MSE) defined as \footnote{This definition is for only one parameter,
  which in our case is the redshift. For a general expression, one
  would also sum over the different parameters.}
\be
L = \frac{1}{N} \sum_{k=1}^{N} (z_{\text{pred},k} - z_{\text{ref},k})^2 \, .
\label{eq:MSE}
\ee
After completing the forward propagation and computing the loss for the batch,
the information is propagated to the weights and biases (back
propagation) that are then modified using a stochastic gradient
descent algorithm with a momentum of 0.9. This procedure is repeated for all batches in the
training set and a total training loss for this epoch is
thus obtained. Afterwards, the loss is computed within the validation set
to determine the improvement of the network, which concludes the epoch.

We perform a so-called cross validation to minimize bias in the
validation set, which comprises training the NN on the training set and
using the validation set to validate the performance after each epoch as
described above. These steps are repeated by exchanging the validation
set within the training set, such that we have with our splitting five
cross-validation runs. In the end, the network is trained on training
and validation set together and terminated at the best epoch of all
cross validation runs. The best epoch is defined as the epoch with the
minimal average validation loss. This network is then applied to the
test set, which contains data the network has never seen before.

%\FloatBarrier
\section{Main Redshift Network NetZ$_\text{main}$}
\label{sec:results}

\begin{figure}[ht!]
  \centering
  \figuretitle{Network NetZ$_\text{main}$ trained on all galaxies with $0<z_\text{ref} \lesssim 4$}
\includegraphics[angle = 0, trim=0 0 0 0, clip, width=1.0\columnwidth]{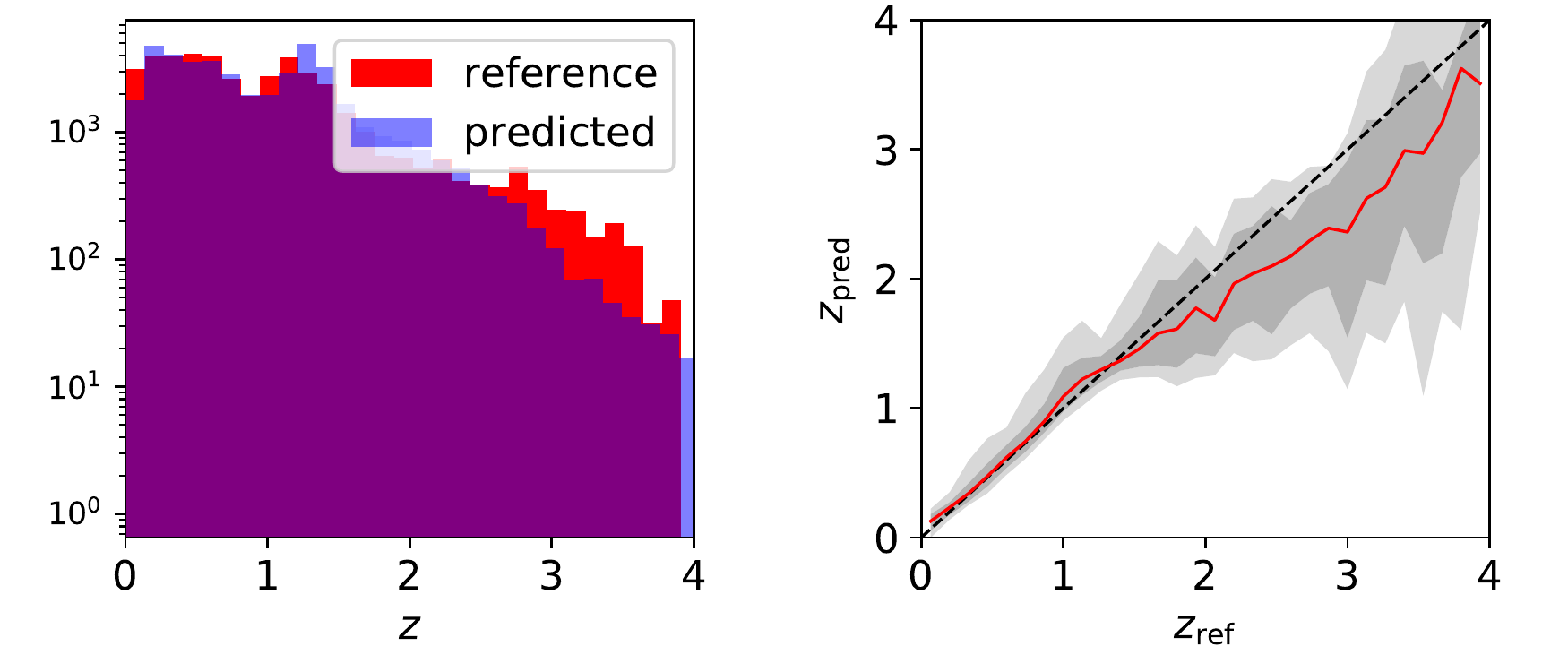}
% this plot is based on test set of Network_0340 as NetZ_main
\caption{Performance of the final network on the test set T. On the
  left hand side, histograms of the reference and predicted redshift
  distributions are shown in red and blue, respectively. On the right
  hand side, a 1:1 comparison of reference and
  predicted redshifts is plotted. The red line shows the median
  predicted redshift per bin and the gray bands the 1$\sigma$ and
  2$\sigma$ confidence levels. While the red line follows the black
  dashed reference line for low redshift very nicely,
  NetZ$_\text{main}$ tends to underpredict the high end.}
\label{fig:comparison}
\end{figure}

In this section, we present our main network NetZ$_\text{main}$ which is trained
in the full redshift range ($0<z \lesssim 4$). We find that this CNN is overall very precise in
predicting redshifts. Figure \ref{fig:comparison} shows a comparison
of our final network predictions $z_\text{pred}$ to the reference
redshifts $z_\text{ref}$ of the test set T. In detail, the left panel
of this plot shows a histogram of the reference redshifts (red) and
predicted values (blue). On the right panel, a 1:1 comparison of
reference and predicted redshifts is plotted. The red line shows the
median and the gray bands the 1$\sigma$ and 2$\sigma$ confidence
levels.

While the network performance is good in the redshift range between 0
and $\sim$2, the network starts to underestimate the higher
redshifts. This is understandable as the network is trained on many
more images in the lower reshift range as we can see directly from the
histogram. The reason is the limited amount of available
training data (reference redshifts) above $z\sim 2$. Moreover, these
distant galaxies are typically faint and small in extent, which
complicates the learning procss with regard to  their morphological
features. 

As described in Sect.~\ref{sec:network}, we use cross-validation and train always over
300 epochs. We do not see much overfitting from the loss curve, where
overfitting means that after a certain number of epochs the network
learns to predict the redshifts better for the training set than for
the validation set. Based on our testing of different hyper-parameters such as batch size
or learning rate, the best moment for terminating the training of
NetZ$_\text{main}$ is at epoch 135 with a loss of 0.1107 according to
the loss function $L$. This network has a learning rate of 0.0005 and
a batch size of 128. We also tested drop-out, which means to ignore
during each training epoch a new random set of neurons. This can help to
reduce overfitting and balance the importance of the neurons in the
network. We carried out a test using a dropout rate of 0.5 between the FC layers, but it turned out that drop-out was not necessary for this network.

% originally in sec "training data"
We tested the network performance by
varying the masking, such as the deblending threshold and the kernel for the smoothing. The
difference of $\lesssim 0.01$ in the predictions is small compared to
the typical photo-$z$ uncertainty (as we see in the scatter of Figure
\ref{fig:comparison}). % Also, if we use not extracted galaxy images the network is able to predict the redshift within acceptable uncertainty.
This network stability is important in case the extraction is not done
perfectly as planned and done for the training. The masking
is done in the exact same way for the newly predicted photo-$z$ values
as for training and testing.

 It turns out that the network predicts similar
but slightly different values for the augmented images, which shows
that the network does not identify the rotated or mirrored images as
duplicates. The possibility to use such data augmentation and hence
boost the performance at high redshifts is a major strength of NetZ.

\begin{figure}[ht!]
  \centering
  \includegraphics[angle = 0, trim=0 0 0 0, clip, width=\columnwidth]{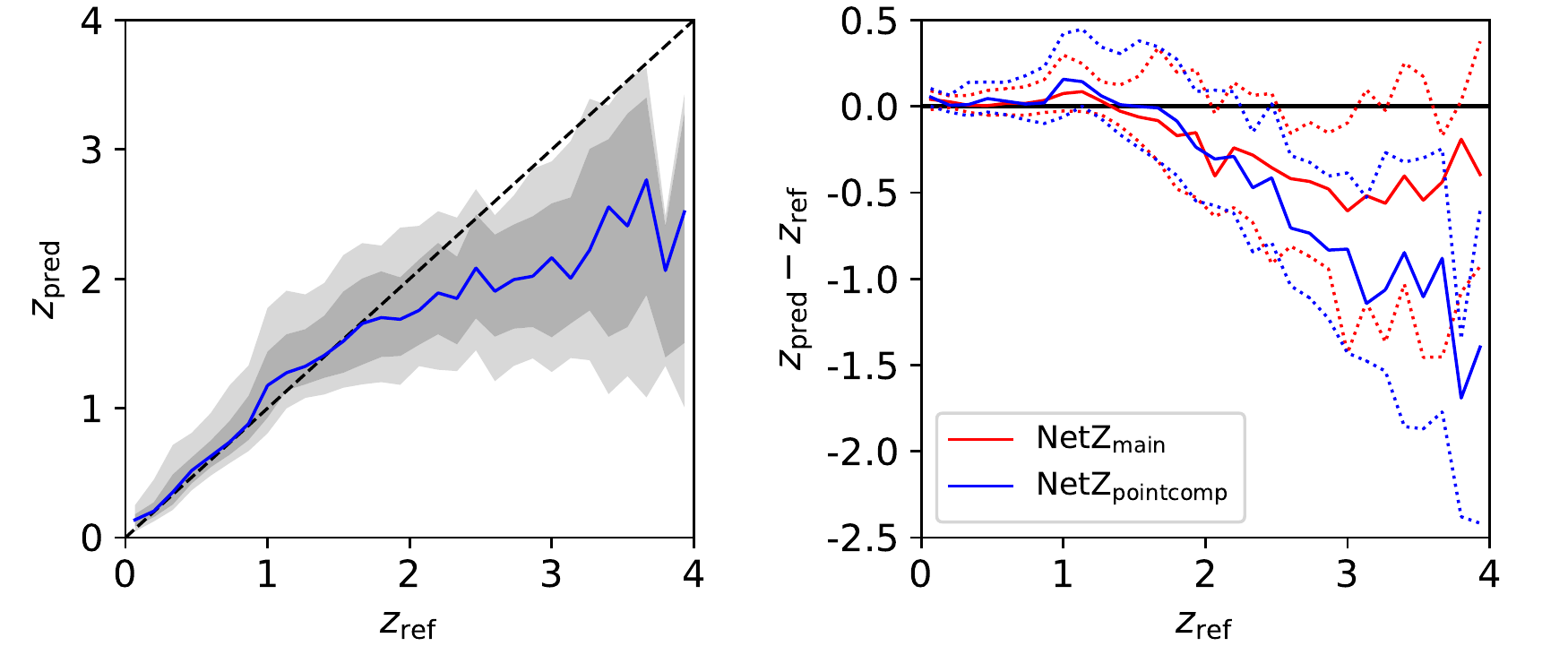}\\
  \caption{Performance of the network trained on images of point-like sources in place of galaxies (blue) with 1 and 2 $\sigma$ ranges on the left (gray), and as a comparison to NetZ$_\text{main}$ (red) on the right panel with 1$\sigma$ ranges (dotted). We directly see that the original galaxy images and thus their morphological information improve the network.}
\label{fig:PSFtest}
\end{figure}

As a further test, we replaced the image cutouts of the galaxies with point-like sources using the corresponding PSF images and scaling them to the correct magnitudes. This way, the images contain only the information available from the catalogs (as used in typical photo-z methods) but exclude any morphological information such as the galaxy shapes from the real cutouts. We tested a few different hyper-parameter combinations by varying the learning rate and also the number of convolutional layers, but with the result of worse performance in predicting redshifts, as shown in Figure~\ref{fig:PSFtest}. For more detail, we show on the left panel the performance of the network trained on the point-like images in analogy to the right panel of Figure~\ref{fig:comparison}; and on the right panel, we make the direct comparison between the network trained on the correct image cutouts (red) and the network trained on images of point-like sources (blue). We can directly see the smaller scatter in the predicted redshift when using the correct cutouts, especially on the high-redshift range even with our use of data augmentation in this redshift range. This test shows the importance of the morphological information for this method and that it contributes significantly to the robustness of the photo-$z$ predictions.

%FloatBarrier
\section{Comparison of NetZ$_\text{main}$ to other photo-$z$ methods}
\label{sec:comparison}

\subsection{Detailed comparison to HSC method DEmP}
\label{sec:comparison:DEmP}

Since there are already different photo-$z$ methods developed and
applied to the HSC data, we show  a comparison here. It is very
important to use the same data set for a fair comparison. Thus, we
can only make a comparison with the DEmP \citep{hsieh14} method, where we have a
predicted photo-$z$ value for each galaxy within our test set T and using identical training and validation sets as for NetZ$_\text{main}$, without data augmentation -- since DEmP also relies
on the reference distribution. DEmP is one of the best-performing
methods from the HSC photo-$z$ team \citep{tanaka18, nishizawa20} and,
thus, it stands as a good performance benchmark. DEmP is a hybrid photo-$z$ method by
combining polynomial fitting and a N-nearest neighbor method based on
photometric values on a catalog level. Therefore, the input data are
totally different from those of NetZ, which is based on the pixelated
image cutouts of the galaxies.

For the comparison, we adopted three
quantities from the HSC photo-$z$ papers \citep{tanaka18, nishizawa20},
which are defined as follows for each redshift bin:

\begin{figure*}[t!]
  \centering
  \includegraphics[angle = 0, trim=  70  35  50  50, clip, width=2\columnwidth]{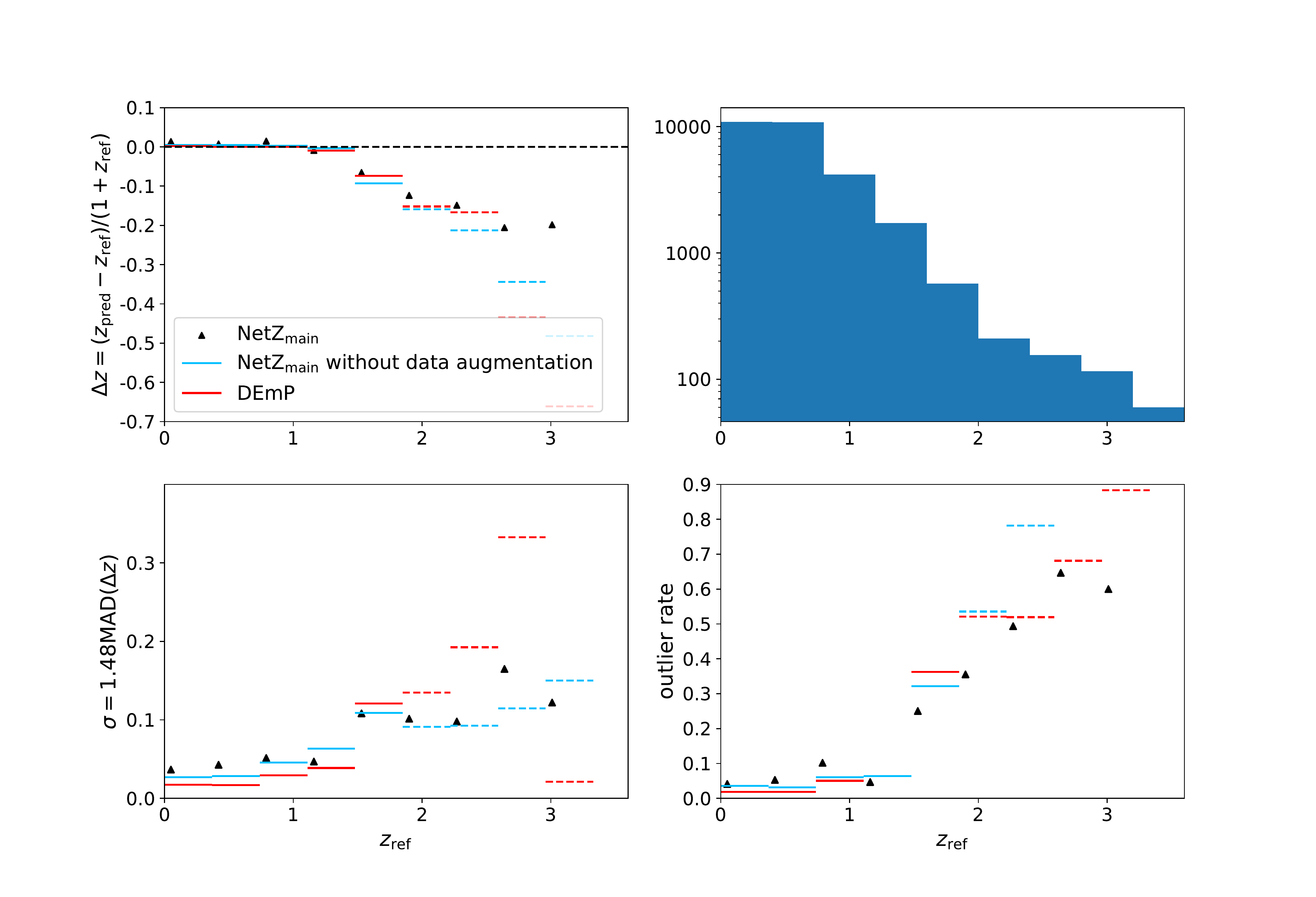}
% this plot is based on test set of Network_0340 as NetZ_main, test set of DEmP, and Network_0310 as NetZ w/o data augmentation. Script is
% make_plot_HSC_comparison_oneFig.py
\caption{NetZ$_\text{main}$ (black points) performance in terms of
  bias (top-left panel), dispersion (bottom-left panel), and outlier
  rate (bottom-right panel) as functions of the reference redshift
  $z_{\text{ref}}$ in comparison to DEmP (red bars).
  Definitions of bias, dispersion and outlier rate are given in
  Eqs.~\ref{eq:bias}-\ref{eq:outliers}.  We show also with blue bars
  the results from a network where we do not use data augmentation to
  increase the number of high-$z$ galaxies.  The values in dashed bars are
  based on limited number of galaxies.  The histogram in the top-right
  panel shows the number of galaxies as a function of redshift in the
  test set T used for the comparison.  NetZ performs substantially
  better than DEmP at $z_{\rm ref} \gtrsim 2$, with smaller bias,
  lower dispersion and lower outlier rate, by up to a factor of 2.}
\label{fig:comparison_plot_HSC}
\end{figure*}

\bea
{\bf Bias:} & \text{Median} \, (\Delta z_i) = \text{Median} \, \left(\frac{z_{\text{pred},i}-z_{\text{ref},i}}{1+z_{\text{ref},i}} \right) \label{eq:bias},\\
{\bf Dispersion:} & \sigma = 1.48 \times \text{MAD}(\Delta z_i) = \nonumber \label{eq:dispersion}\\
         & = 1.48 \times \text{Median} \, \left( |\Delta z_i -\text{Median}({\Delta z_i})| \right), \\
{\bf Outlier ~rate:} & f_\text{outlier} = \frac{N( |\Delta z_i | > 0.15)}{N_\text{bin}} \label{eq:outliers}
.\eea

\noindent where $i$ denotes the $i^{\rm th}$ galaxy in the redshift bin,
$z_{\text{pred}}$ the predicted photometric redshift, $z_{\text{ref}}$
the reference redshift, $N$ the number of galaxies satisfying the
specified condition, and $N_\text{bin}$ the total number of galaxies
in the bin.  The dispersion is defined using the median absolute
deviation (MAD), as expressed above. The multiplication factor comes
from statistics and is the relation factor for normally distributed
data between MAD and the standard deviation \citep{rousseeuw93}.

The
comparison is shown in Figure \ref{fig:comparison_plot_HSC}, with black
triangles showing the performance of NetZ$_\text{main}$ and red bars showing the
performance of DEmP. Since we use data augmentation (rotation and
mirroring of images; see Sect.~\ref{sec:data} for details) to
increase the number of high-$z$ galaxies, which is not possible for DEmP,
we also trained  a network without data augmentation -- we show both in
Figure \ref{fig:comparison_plot_HSC} for comparison. For the range
$z_{\rm ref}\lesssim1.5$, the performances of both methods are very
good especially for the bias, with DEmP performing slightly better
than NetZ. If we compare the range $z_{\rm ref} \gtrsim 1.5$, the
performance of both methods decreases, but NetZ with data augmentation  now performs noticeably
better than DEmP.

A decrease in performance in the redshift range
around $z \approx 2$ is expected, as the used filter set $grizy$ does
not cover the prominent $4000\AA$ break but, in contrast to the other
methods presented in \citet{tanaka18}, NetZ and partly DEmP can at least
break the degeneracy between low redshifts ($z<0.5$) and the
redshift range around 3 to 4, which is not the case for Mizuki from \citet{nishizawa20} and several methods from \citet{tanaka18} that used HSC images with similar reference redshifts. A more detailed and direct comparison is difficult since the predicted redshifts from these methods are not publicly available for our whole test set; moreover, some of the galaxies in our test set could be in the training data of these methods, which would artificially improve their performance.

The very low dispersion of DEmP in the highest-redshift bin comes from
DEmP underestimating consistently most of the redshifts, and hence the
outlier rate is large. Although the outlier rate is high in the range
$z_{\text{ref}} \gtrsim 2$ in general, the performance is primarily
limited by the number of existing reference redshifts in this
range.
While
DEmP is developed and tested with a big enough training sample also for higher
redshifts, we used here for DEmP the exact same data set as for NetZ
for a fair comparison. Since there is no sufficient training sample at $z_\text{ref}>2$ for both NetZ without data 
augmentation and DEmP, we plot these values in dashed because they are less reliable. It is nonetheless
encouraging to see the significant reduction in the bias, dispersion, and
outlier rate of NetZ with data augmentation for the high-$z$ range, up to a factor of 2
relative to DEmP, thanks to the use of the spatial information from the
galaxy images in addition to photometry. Especially for upcoming
surveys such as LSST, which will provide relatively deep images, it is
important to have methods prepared and tested on the higher redshift
range.

%% originally fig:comparison_plot_HSC here

\begin{figure}[ht!]
  \centering
  \includegraphics[angle = 0, trim=0 0 0 0, clip, width=\columnwidth]{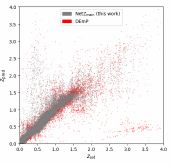}\\
  \caption{Network performance as scatter plot comparing the predicted
    with the reference redshifts for NetZ$_\text{main}$ (this work)
    and DEmP \citep{hsieh14}. The scatter looks overall comparable at
    $z_{\rm ref}\lesssim2$, while NetZ$_\text{main}$ does not contain
    the catastrophic outliers at $z_\text{ref} \sim 3$ and
    $z_\text{pred} \sim 0.5$ that DEmP has.}
\label{fig:scatterplot}
\end{figure}

As a further comparison, we show in Figure \ref{fig:scatterplot} a
scatter plot of $z_\text{pred}$ versus $z_\text{ref}$ for DEmP
\citep{hsieh14} and our neural network NetZ$_\text{main}$ with data augmentation. From this plot we can again  see 
the good performance for the low-$z$ range, where we note that the
number of outliers from NetZ is negligible compared to the number of
galaxies in the bins, which is also evident in the outlier rate. If we
assume that all catastrophic outliers for $z_\text{ref}<1.5$ are
misplaced at high redshift, which is very conservative, then $>$ 77\%
of the galaxies predicted to be at $z_{\rm pred} > 1.5 $ are actually
at $z_{\rm ref} > 1.5$. In the high-$z$ range, NetZ tends to predict too
low redshifts, but it does not have the cluster of catastrophic
outliers at $z_\text{pred} \sim 0.5$ and $z_{\text{ref}}$ between 3
and 3.5  that DEmP does and this is due to the Lyman-break or Balmer-break
misclassification. Even for the galaxies where the NetZ
redshifts are classified as outliers, these redshifts are closer to
the true redshift than for DEmP. The outlier rate for NetZ is
dominated by blue-star-forming galaxies and galaxies with a small
spatial extent (covering $\approx 20-30$ pixels) that provide little
information for the CNN to extract features. We therefore note that
galaxies covering a small number of pixels are more prone to be
catastrophic outliers in their redshifts and should be treated with
caution.

\subsection{Photo-$z$ with morphological information}
\label{sec:comparison:morpho}

The studies presented by, for instance, \citet{soo18} and 
\citet{wilson20} aim to include morphological information of galaxies
to improve photo-$z$ estimations.  
%very different to NetZ$_\text{main}$ in that a conclusion of the
%improvement through the additional morphological information is
%difficult. 
In particular, \citet{wilson20} make use of optical and near infrared
observations, some of which are obtained with the Hubble Space Telescope (HST). Therefore the considered data
cover a wider wavelength range and are additionally of better spatial
resolution than our ground based HSC images. However, \citet{wilson20}
consider only photometric measurements and four morphological
measurements (half-light radius, concentration, asymmetry, and
smoothness), rather than the pixels directly. By working directly with
the image pixels in our CNN, we use the maximal amount of information and we are
independent of the pipelines and uncertainties when extracting morphological
measurements. Moreover, \citet{wilson20} limit the range to $0 < z < 2$,
which makes the network not directly applicable to deep imaging surveys, especially since we are focusing on the
high-redshift range. As we show in the next section, we also obtain very
good results within a limited range. 
With these differences in the assumptions and data sets, it
is difficult to directly compare the results of \citet{wilson20} and
NetZ. Nonetheless, if we compare our outlier fraction with $\sim$0.05 
up to $z \sim 1.7$ (see Figure \ref{fig:comparison_plot_HSC}) to
that from \citet{wilson20}, which is called OLF, with $\sim$0.1-0.2 up
to $z=2$ (see Tables 2 and 3 of \citet{wilson20}), NetZ yields a good
improvement. 
%The comparison of the bias is difficult as their Figures 2
%and 3 show the values for each individual point while the bias
%is from us defined as their median. 
While \citet{soo18} and \citet{wilson20} find that morphological measurements do not provide a notable improvement in photo-$z$ predictions when compared to using only multi-band photometric measurements, our NetZ results (Figure~\ref{fig:PSFtest}) show that the pixels in the image cutouts that contain morphological information are useful. This suggests that a promising avenue for future developments of photo-$z$ methods is to combine photometric measurements (as typically used for current photo-$z$ methods) with direct image cutouts (as used for NetZ) instead of morphological measurements.

\subsection{Photo-$z$ estimates for LSST}
\label{sec:comparison:LSST}

\citet{schmidt20} present a collection of different photo-$z$ methods tested on LSST mock data. In particular, they compare 12 different codes, of which three methods are based on template fitting (BPZ, \citet{benitez00}; EAZY, \citet{brammer08}; LePhare, \citet{arnouts99}), seven are based on machine learning (ANNz2, \citet{sadeh16}; CMN,\citet{graham18}; FlexZBoost, \citet{izbicki16}; GPz, \citet{almosallam16b}; METAPhoR, \citet{cavuoti17}; SkyNet, \citet{Graff14}; TPZ, \citet{carrascoKind13}), one is a hybrid method \citep[Delight,][]{leistedt17}, and one is a pathological photo-$z$ PDF estimator method \citep[trainZ,][]{schmidt20}. The last method trainZ is designed to serve as an experimental control, and not a competitive photo-z PDF method. It assigns to each galaxy set a photo-$z$ PDF by effectively performing a k-nearest neighbor procedure. As training data, they use $<10^7$ LSST like mock data limited to $0<z<2$ and an $i$ band magnitude limit of 25.3 to match the LSST gold sample \citep[for further details see][]{schmidt20}. The main advantage of these methods in \citet{schmidt20} compared to the current version of NetZ is the probability density function estimates, whereas NetZ does not require photo-$z$ pre-selection and shows a good performance over a broader redshift range ($0<z<4$). Based on the different redshift range and data sets, a detailed and fair comparison is not possible. If we compare Figure~\ref{fig:comparison} to Figure~B1 of \citet{schmidt20} quantitatively, we see an overall similar performance, but most of the LSST methods have a cluster of outliers at $z_\text{ref} \sim 0.5$ and $z_\text{ref} \sim 1.7$ which we do not see with NetZ. The kink at $z_\text{ref} \sim 1.7$ might be related to the drop of data points and an edge effect near the end of the assumed range since we observe a similar effect with NetZ for higher redshifts ($z_\text{ref} \sim 3$). Comparing the machine learning methods is difficult as well. The network architectures, as with nearest-neighbour algorithms, random forests, prediction trees, or sparse Gaussian processes, which are presented in \citet{schmidt20}, are simply too different from the image-based CNN we present with NetZ.

%\FloatBarrier
\section{Limited-range and LRG-only redshift network}
\label{sec:specific}

\begin{figure}[ht!]
  \centering
  \figuretitle{Network NetZ$_\text{lowz}$ trained on all galaxies with $0 < z_\text{ref} \leq 1$}
\includegraphics[angle = 0, trim=0 5 25 0, clip, width=1.0\columnwidth]{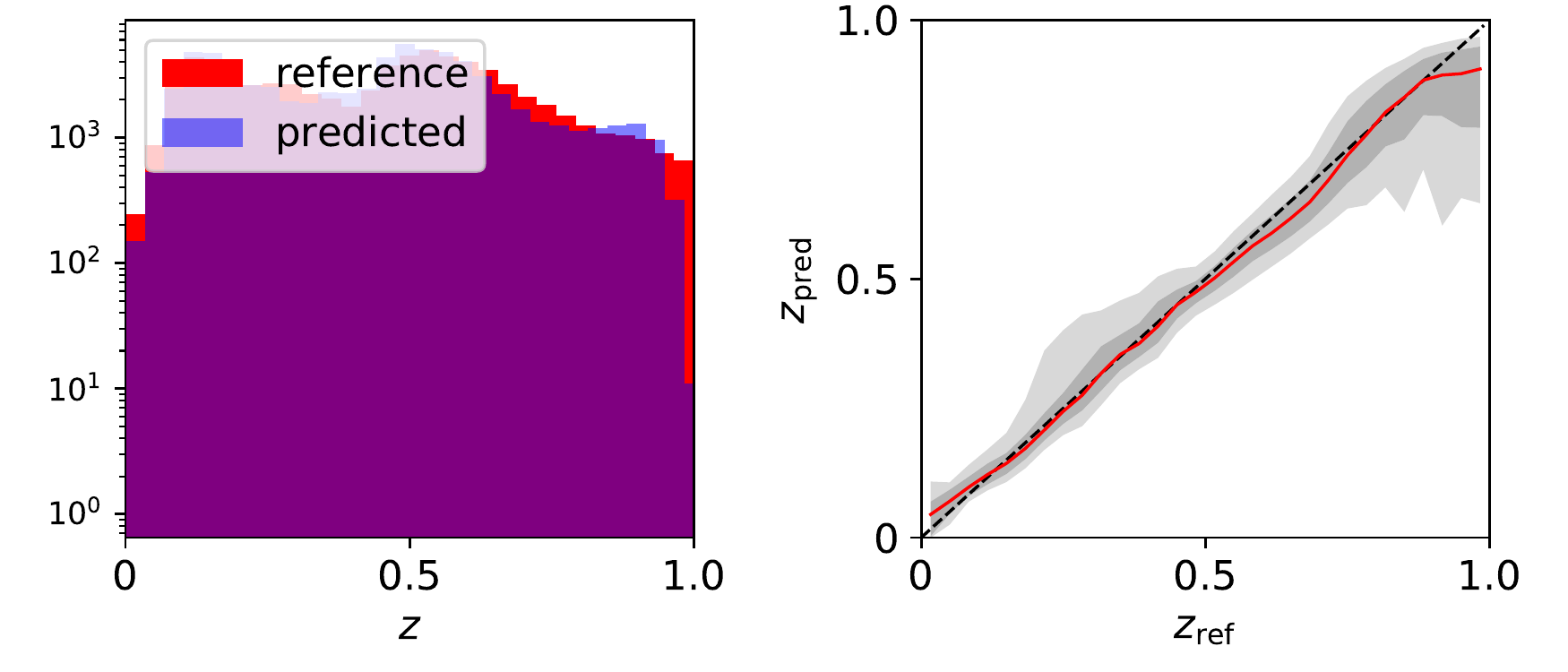}
\caption{Performance of the network NetZ$_\text{lowz}$ trained on all types of galaxies
   in the range $0 < z_\text{ref} \leq 1$. On the left hand side,
  histograms of the redshift distributions are shown, in red the
  distribution of the reference redshifts used to train the network
  (ground truth) and in blue the predicted redshift distribution.  On
  the right panel, a 1:1 comparison of reference and predicted
  redshifts is plotted. The red lines show the median and the gray
  bands show the 1$\sigma$ and 2$\sigma$ confidence levels.}
\label{fig:comparisonlowz}
\end{figure}

During our testing, we found we could obtain substantial improvement by restricting
the redshift range. We explored, for instance, networks with redshift
ranges limited to $0<z<1$ and $1<z<2$, but not to higher redshift
intervals, due to the limitations in available reference redshifts for
$z>2$. Limiting to $0<z<1$ is also done in several other works
\citep[e.g.,][]{hoyle16, pasquet19, campagne20}. To benefit from
these refined networks in any practical application, we would first
need to predict the correct redshift range and then these
networks could be used in a specified range. We also considered combining
multiple networks and iteratively refining the photo-$z$ predictions,
that is, start with NetZ$_\text{main}$ to predict $z_\text{pred}$ and then,
based on the value of $z_\text{pred}$, we could  subsequently apply a network
that is trained in a narrower range around $z_\text{pred}$ to refine
the $z_\text{pred}$ estimate. However, we find that outliers from
NetZ$_\text{main}$ limit the gain we can achieve in refining
$z_\text{pred}$. A practical possibility to use a redshift network for
the lower-redshift end of the distribution would be to restrict the
sample by the galaxy brightness. If we restrict our data set D to
galaxies with an apparent $i$-band AB magnitude brighter than 22, the catalog
includes only 1.3\% objects with $z_\text{ref} \geq 1$ and we miss 12.9\%
of all galaxies from the original set D with $z_\text{ref} \leq 1$. For the
training of NetZ$_\text{lowz}$ itself we limit only to a narrow redshift range but not in
magnitude. The performance of NetZ$_\text{lowz}$ is shown in Figure~\ref{fig:comparisonlowz}, on the left a histogram of the reference (red) and predicted (blue) redshifts. The distribution of the predicted redshift follows that of the reference redshift very well. On the right side, we show a 1:1 correlation plot, with the median as a red line and in gray the $1\sigma$ and $2\sigma$ areas. If we compare the two (Figure~\ref{fig:comparison}), we can see that NetZ$_\text{lowz}$ performs significantly better than NetZ$_\text{main}$, as expected.

\begin{figure}[ht!]
\centering
\includegraphics[angle = 0, trim=0 0 0 0, clip, width=\columnwidth]{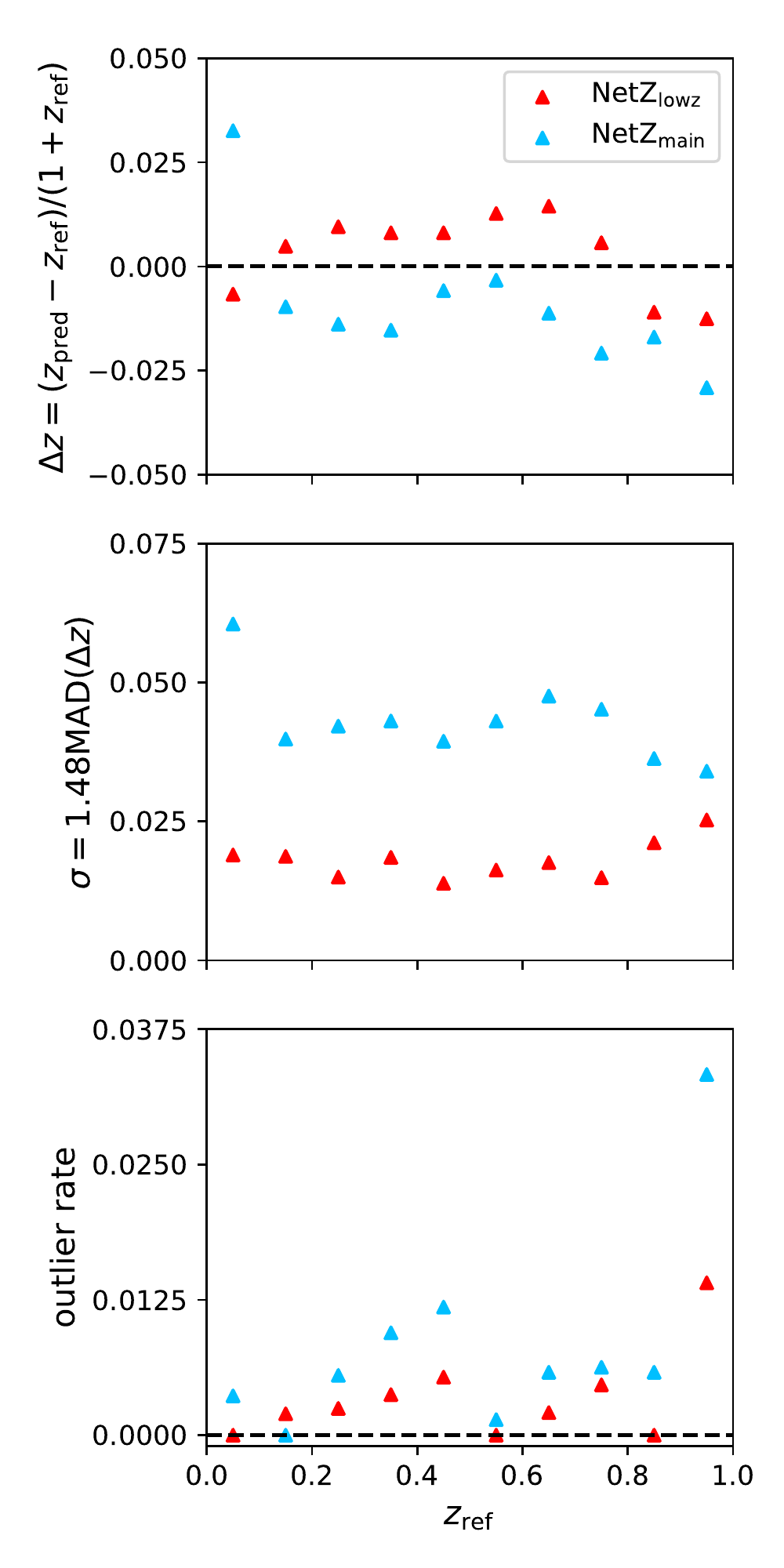}
% make_plot_HSC_comparison_LowZ.py on 2020-11-17
% this plot is based on network_0191 as NetZ_lowz, but only those galaxies with i-mag >22 of the test set and those in the test set of Network_0340
% compare to Network_0340 , to which we compare as NetZ_main 
\caption{Network performance of NetZ$_\text{lowz}$ compared to NetZ$_\text{main}$ in terms of bias,
  dispersion, and outlier rate (see
  eq. \ref{eq:bias}-\ref{eq:outliers}) as functions of the reference
  redshift $z_{\text{ref}}$. For this comparison, we use the overlap between both test sets and only galaxies with an $i$-band magnitude brighter than 22 as NetZ$_\text{lowz}$ would be applied only to them. }
\label{fig:comparisonNetZLowz}
\end{figure}

\begin{figure}[ht!]
\centering
\includegraphics[angle = 0, trim=0 0 0 0, clip, width=\columnwidth]{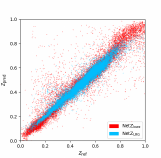}
\caption{Predicted redshifts $z_\text{pred}$ against the reference
  redshifts $z_\text{ref}$ for the networks NetZ$_\text{LRG}$ and
  NetZ$_\text{lowz}$ of their test set. We see directly a lower outlier rate for
  NetZ$_\text{LRG}$ than NetZ$_\text{lowz}$.}
\label{fig:scatterLRG}
\end{figure}

We further show in Figure~\ref{fig:comparisonNetZLowz} the bias, dispersion, and outlier rate for NetZ$_\text{lowz}$ (red). If we compare this performance to NetZ$_\text{main}$ applied to the same galaxies for a fair comparison (blue), we find a good improvement in the bias and, with a factor of $\sim 2$ reduction, in the dispersion. Only the outlier rate is comparable. If we compare the performance of
NetZ$_\text{main}$ on the full test set that of  the network, we still see an improvement for the network
NetZ$_\text{lowz}$ without the $i$ magnitude limitation. This confirms
that the improvement is related to the network range. A scatter plot of NetZ$_\text{lowz}$ is shown in Figure~\ref{fig:scatterLRG}.

\begin{figure}[ht!]
  \centering
  \figuretitle{Network NetZ$_\text{LRG}$ trained on LRGs-only with $0 < z_\text{ref} \leq 1$}
\includegraphics[angle = 0, trim=0 5 25 0, clip, width=1.0\columnwidth]{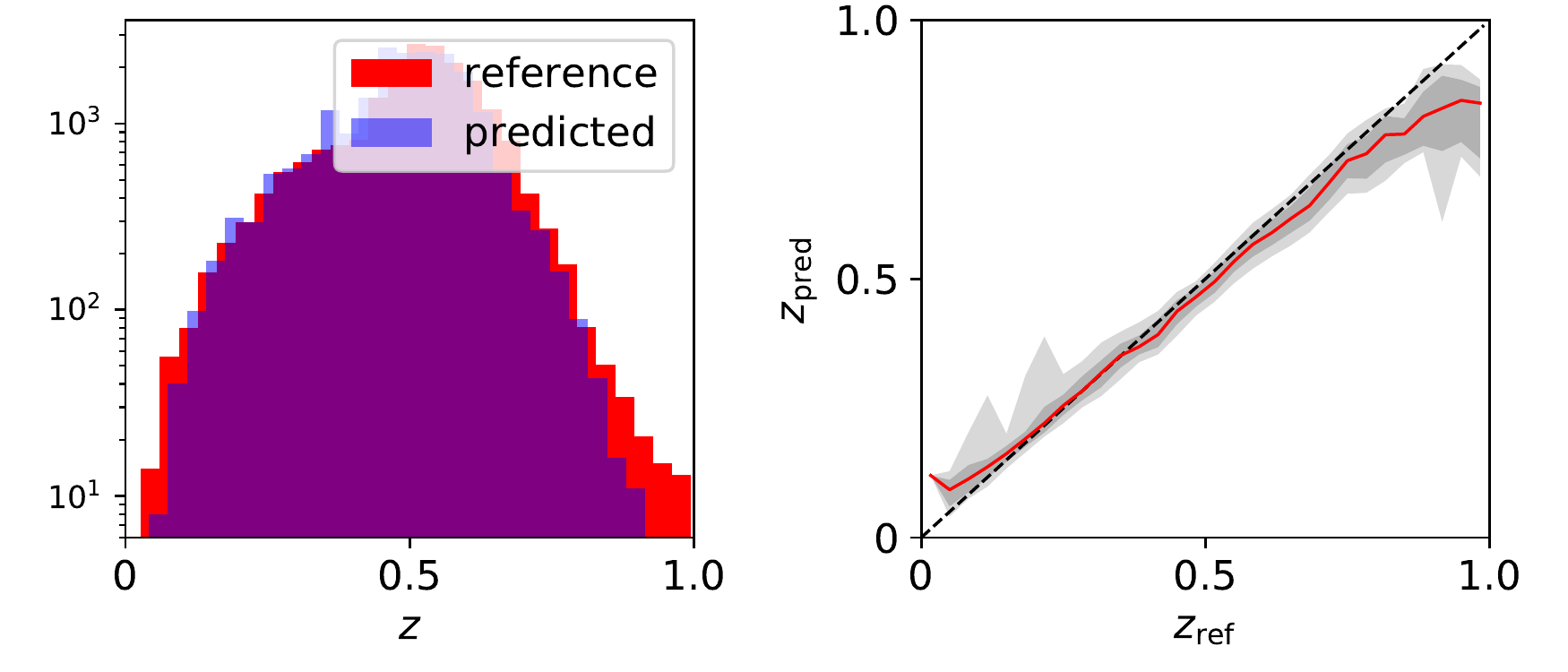}
\caption{Performance of the network NetZ$_\text{LRG}$ (bottom) trained on LRGs only. On the left-hand side,
  histograms of the redshift distributions are shown, in red we show the
  distribution of the reference redshifts used to train the network
  (ground truth) and in blue the predicted redshift distribution.  On
  the right panel, a 1:1 comparison of reference and predicted
  redshifts is plotted. The red lines show the median and the gray
  bands the 1$\sigma$ and 2$\sigma$ confidence levels.
}
\label{fig:comparisonLRGs}
\end{figure}

Instead of applying networks trained for specific redshift ranges,
which is difficult to do in practice, we can consider specific classes
of galaxies that can be selected a priori, such as LRGs. Therefore we investigate a redshift estimation
network specialized on LRGs that are useful for various studies
including strong lensing and baryon acoustic oscillations. Since
nearly all LRGs out of our reference sample have $z_\text{pred}< 1$,
we show here the network performance of NetZ$_\text{LRG}$ in
comparison to the network NetZ$_\text{lowz}$ trained on all galaxy types. Figure
\ref{fig:comparisonLRGs} shows on the left a histogram and on the
right the 1:1 comparison of $z_\text{ref}$ and $z_\text{pred}$.

\begin{figure}[ht!]
\centering
\includegraphics[angle = 0, trim=0 0 0 0, clip, width=\columnwidth]{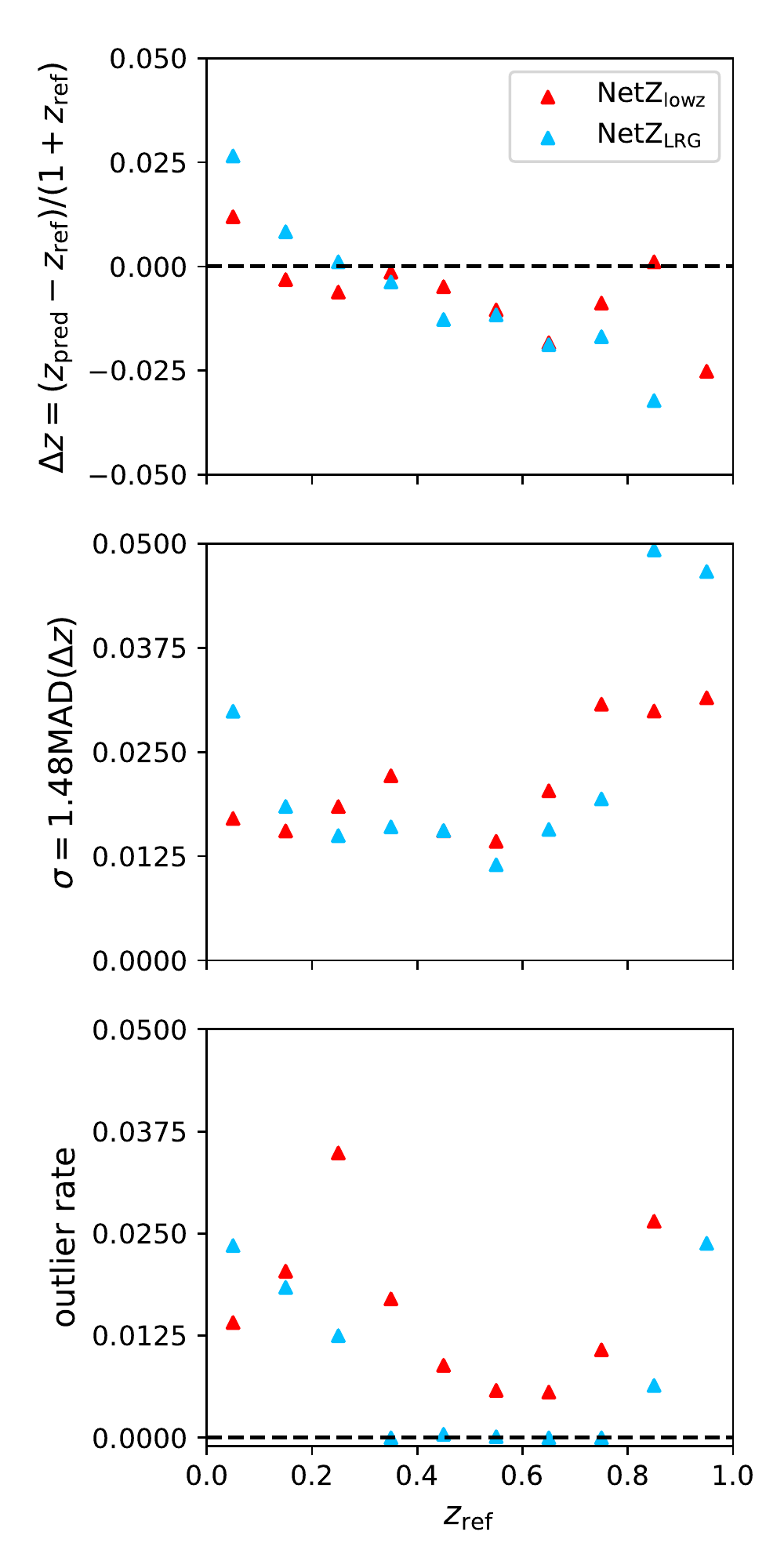}
% this plot compares Network_0199 as NetZ_LRG with
% Network_0191 as NetZ_lowz but now applied only to LRGs
\caption{Network performance of NetZ$_\text{LRG}$ compared to
  NetZ$_\text{lowz}$ applied to LRGs only in terms of bias, outlier
  rate, and dispersion (see eq. \ref{eq:bias}-\ref{eq:outliers}) as
  functions of the reference redshift, $z_{\text{ref}}$.}
\label{fig:comparisonLRGtoALL}
\end{figure}

We show further the bias, dispersion, and outlier rate (defined in eq
\ref{eq:bias}-\ref{eq:outliers}) in Figure
\ref{fig:comparisonLRGtoALL}. The network NetZ$_\text{LRG}$ performs
better in most redshift bins. Finally, in
Figure \ref{fig:scatterLRG} we show a scatter plot of this network without magnitude limitation in comparison to NetZ$_\text{lowz}$. From this we can see again the
redshift limits of the LRG sample and also the good improvement.

Both networks NetZ$_\text{LRG}$ and NetZ$_\text{lowz}$ show that
photo-$z$ for subsamples of galaxies does overall better than the main
network NetZ$_\text{main}$ that is trained on all galaxies. Therefore,
for specific subsamples, it would be beneficial to train a CNN
specific to that sample.

%\FloatBarrier
\section{Summary and conclusions}
\label{sec:summary}
% The convolutional neural network with the architecture described in
% Section \ref{sec:network} is able to predict very accurate the
% redshift based on the images of the galaxy. The input for the
% network are the images of the galaxy in five different filter,
% \textit{grizy}. For training the network we are using data from the
% HSC survey, but the underlying concept should work also for
% different/better image qualities. Since the image quality of LSST is
% very imilar to that from HSC, the here presented network should also
% be able to handle those images.

With current and upcoming imaging surveys, we anticipate that billions of
galaxies will be the subject of observations, while just a small fraction of them will have
spectroscopically confirmed redshifts. Therefore, it is necessary to
have tools to obtain good photometric redshifts, especially for the
higher redshift range as the upcoming surveys will provide deeper
images where previous photo-$z$ methods have strong limitations. With
the success of ML and especially CNNs in image processing, we
investigated a new CNN based technique to estimate the photo-$z$ of a
galaxy. The method is very general; it accepts directly cutouts of the
observed images and predicts the corresponding redshift. Therefore it is directly applicable to all HSC cutouts after applying simple cuts on the Kron radius and $i$ band magnitude observables.

For
training the network and testing the performance, we carry out a comparison with to
reference redshifts from various, mostly magnitude-limited surveys. In this paper,
we focus on HSC data with a pixel size of $0.168\arcsec$ and use the
available five filters \textit{grizy}, which are also part of the upcoming LSST\footnote{LSST has in addition u-band observations.}. In principle, it is also possible to include additional filters, such as the near-infrared (NIR) range from the same or a different telescope which would improve the performance even more, as shown by \citet{gomes18}, for the low redshift range ($z \lesssim 0.6$). The only constraint from the CNN is the constant pixel size over all different filters. Since NIR images have typically larger pixel sizes, an interpolation and resampling to the same pixel resolution as the optical images would be necessary. What remains to be seen is how much NetZ could benefit from such additional filters, especially in the high-z range, and this would need to be tested.  In addition, NetZ could be trained on additional Euclid images that are high-resolution from space in the visible and infrared range. Even without combining Euclid with ground-based images, our photo-z estimates from HSC are useful for Euclid given the overlap in the footprints of HSC and Euclid.

With our trained network on HSC images, we find an overall very good performance of the network with a
1$\sigma$ uncertainty of 0.12 averaged over all galaxies from the
whole redshift range. Our CNN provides a point estimate for each
galaxy with uncertainties adopted from the scatter in each redshift bin of the test set. Based
on the amount of available data, the network performs better in the
redshift range below $z = 2$. In the range above $z = 2$, we are using, as a way of gaining an advantage over state-of-the art methods like DEmP, data
augmentation by rotating and mirroring the images. While the bias for DEmP and NetZ$_\text{main}$ as well as the dispersion for DEmP increases significantly in this range, with NetZ$_\text{main}$ we obtain by using data augmentation similar values as for the lower redshift range. We also obtain  better outlier rates for the highest redshift bins by using data augmentation but the improvement is less pronounced. In
particular, NetZ does not under-predict the redshifts of galaxies with
$z_\text{ref} \sim 3 - 3.5$ by as much as DEmP and other methods due
to the Lyman and Balmer break misclassification. The main limitations
that all photo-$z$ methods face when predicting redshifts for
distant galaxies is the low number of reference redshifts. In our case,
the number drops by a factor of around 1000 compared to the range where
$z<2$. Therefore, using the image cutouts gives a good advantage as we can use data augmentation by rotating and mirroring the images. The effect is impressive as one can see in Fig~\ref{fig:comparison_plot_HSC}. Since this is not possible for other photo-$z$ methods, several of them focus only on the
lower redshift range $z<1$ or even lower \citep[e.g.,][]{hoyle16,
  pasquet19, campagne20}. If we also limit the redshift range to
$z<1$, we find a substantial improvement in our network`s
performance. 

In cases where we set our focus on a specific galaxy type like LRGs, we find a
further improvement with regard to the network. This is understandable as the
network can learn better the specific features of this galaxy
type. Based on the small number of LRGs with redshift above $z=1$, we
limit the range of NetZ$_\text{LRG}$ to $0<z<1$ and compare it to a
network trained on all galaxy types in the same redshift range for a
fair comparison.

This paper provides a proof of concept for using a CNN for
photo-$z$ estimates. Based on the encouraging results of NetZ
particularly at high redshifts, we propose further investigations
along the lines of combining our CNN with a nearest-neighbor algorithm
or a fully-connected network that ingests catalog-based photometric
quantities \citep[see][]{LealTaixe16}. 
There are several methods, like DEmP and other methods
\citep[e.g.][]{DIsario18, schmidt20}, which provide a probability
distribution function for the redshifts. Further developments of our CNN approach to provide a probability distribution function of the photo-$z$ require more complex networks such as Bayesian neural networks \citep[e.g.,][]{Levasseur+17} or mixture density networks \citep{DIsario18, hatfield20, eriksen20}. While this is beyond the scope of the current paper, such Bayesian or mixture networks are worth exploring.

\begin{figure}[ht!]
  \centering
  \includegraphics[angle = 0, trim=0 0 0 0, clip, width=\columnwidth]{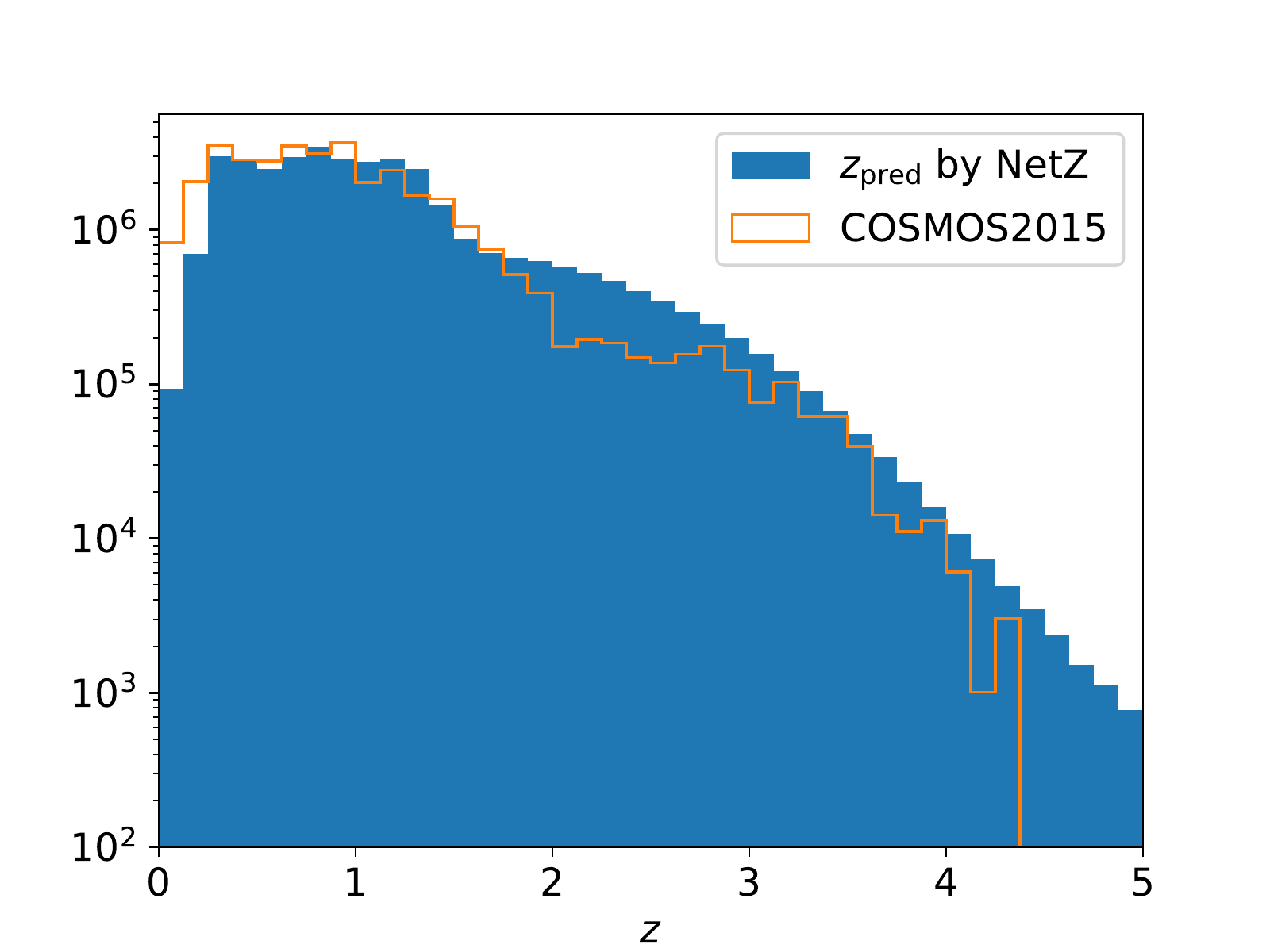}\\
  \caption{Histogram of the newly predicted photo-$z$ values with NetZ based on images of the HSC PDR2 
    (blue filled) and, for comparison, the distribution of COSMOS2015
    \citep{laigle18} scaled by a factor of 1010 to have the same sample
    size (orange open).  The similarity in the two distributions
    shows that NetZ produces a realistic galaxy redshift
    distribution.}
\label{fig:hist_comparison}
\end{figure}

In this work, we show that a very simple convolutional neural
network is able to predict accurate redshifts directly from the
observed galaxy images. NetZ therefore has the advantage of using
maximal information from the intensity pixels in the galaxy images,
rather than relying on photometric or morphological measurements that
could be prone to uncertainties and biases, especially for images of blended
galaxies.  
We ran NetZ$_\text{main}$ on 34,414,686
galaxies from the HSC public data release 2 (PDR2) wide survey and
provide the catalog
here\footnote{The catalog is available at \url{https://www.dropbox.com/sh/grjfo0gkcxsj9n2/AAD-B7D6m7_1i6GGTX0Ionwja?dl=0}}
. We flagged all negative predictions and clear catastrophic outliers ($z_\text{pred}>5$), which are 15,043 and 3,314 objects, respectively, as $-99$. In
Figure \ref{fig:hist_comparison}, we show a histogram of the newly
available photo-$z$ values (blue filled), whose distribution resembles
the magnitude-limited sample of the cleaned COSMOS2015 \citep[][orange
histogram]{laigle18}, which was scaled by a factor of 1010, to have the
same sample size for the purposes of making a direct comparison. This check shows that our NetZ
predictions indeed produce a realistic galaxy redshift distribution
expected for a depth similar to that of LSST. 

As the image quality, depth
and processing of 
HSC and LSST first-year data are expected to be similar (the image processing
pipeline of HSC is a branch of the LSST pipeline), the method we have developed
here will be directly applicable and beneficial to the LSST. The additional
\textit{u}-filter in the LSST will likely further improve photo-$z$ predictions.  
When applying our method to the LSST data, we do not expect to necessarily have to test the network
architecture, however, it is likely that some hyper-parameter combinations would need testing. Since training is more optimally carried out on real images, rather than on mock
images as done, for instance, in \citet[and references therein]{schmidt20}, we suggest that it is optimal to
train a new network on LSST images as soon as they are available.

\FloatBarrier
\begin{acknowledgements}
  We thank Andreas Breitfeld from our IT group for helpful support.\\
  SS, SHS, and RC thank the Max Planck Society for support through the
  Max Planck Research Group for SHS. This project has received funding
  from the European Research Council (ERC) under the European Union’s
  Horizon 2020 research and innovation programme (LENSNOVA: grant
  agreement No 771776).
  \\
  The Hyper Suprime-Cam (HSC) collaboration includes the astronomical
  communities of Japan and Taiwan, and Princeton University. The HSC
  instrumentation and software were developed by the National
  Astronomical Observatory of Japan (NAOJ), the Kavli Institute for
  the Physics and Mathematics of the Universe (Kavli IPMU), the
  University of Tokyo, the High Energy Accelerator Research
  Organization (KEK), the Academia Sinica Institute for Astronomy and
  Astrophysics in Taiwan (ASIAA), and Princeton University. Funding
  was contributed by the FIRST program from Japanese Cabinet Office,
  the Ministry of Education, Culture, Sports, Science and Technology
  (MEXT), the Japan Society for the Promotion of Science (JSPS), Japan
  Science and Technology Agency (JST), the Toray Science Foundation,
  NAOJ, Kavli IPMU, KEK, ASIAA, and Princeton University.
  %\\
  %This paper makes use of software developed for the Legacy Survey of
  %Space and Time. We thank the LSST Project for making their code
  %available as free software at http://dm.lsst.org
  \\
  This paper is based in part on data collected at the Subaru
  Telescope and retrieved from the HSC data archive system, which is
  operated by Subaru Telescope and Astronomy Data Center (ADC) at
  National Astronomical Observatory of Japan. Data analysis was in
  part carried out with the cooperation of Center for Computational
  Astrophysics (CfCA), National Astronomical Observatory of Japan. 
  \\
  We make partly use of the data collected at the Subaru Telescope and
  retrieved from the HSC data archive system, which is operated by
  Subaru Telescope and Astronomy Data Center at National Astronomical
  Observatory of Japan.
  
\end{acknowledgements}

\bibliographystyle{aa}
\bibliography{NetZ}

\end{document}